\documentclass[a4paper,11pt]{article}
\pdfoutput=1
\usepackage{bm}

\usepackage{jcappub} 

\title{\boldmath{Cosmic sign-reversal: non-parametric reconstruction of interacting dark energy with DESI DR2}}

\author[a]{Yun-He Li}

\author[a,b,c,\ast]{and Xin Zhang\note[$\ast$]{Corresponding author.}}
\affiliation[a]{Liaoning Key Laboratory of Cosmology and Astrophysics, College of Sciences, Northeastern University, Shenyang 110819, China}
\affiliation[b]{MOE Key Laboratory of Data Analytics and Optimization for Smart Industry, Northeastern University, Shenyang 110819, China}
\affiliation[c]{National Frontiers Science Center for Industrial Intelligence and Systems Optimization, Northeastern University, Shenyang 110819, China}
\emailAdd{liyunhe@neu.edu.cn}
\emailAdd{zhangxin@neu.edu.cn}

\abstract{A direct interaction between dark energy and dark matter provides a natural and important extension to the standard $\Lambda$CDM cosmology. We perform a non-parametric reconstruction of the vacuum energy ($w=-1$) interacting with cold dark matter using the cosmological data from DESI DR2, Planck CMB, and three SNIa samples (PP, DESY5, and Union3). By discretizing the coupling function $\beta(z)$ into 20 redshift bins and assuming a Gaussian smoothness prior, we reconstruct $\beta(z)$ without assuming any specific parameterization. The mean reconstructed $\beta(z)$ changes sign during cosmic evolution, indicating an energy transfer from cold dark matter to dark energy at early times and a reverse flow at late times. At high redshifts, $\beta(z)$ shows a $\sim 2\sigma$ deviation from $\Lambda$CDM. At low redshifts, the results depend on the SNIa sample: CMB+DESI and CMB+DESI+PP yield $\beta(z)$ consistent with zero within $2\sigma$, while CMB+DESI+DESY5 and CMB+DESI+Union3 prefer negative $\beta$ at $\sim2\sigma$. Both $\chi^2$ tests and Bayesian analyses favor the $\beta(z)$ model, with CMB+DESI DR2+DESY5 showing the most significant support through the largest improvement in goodness of fit ($\Delta\chi^2_{\rm MAP}=-17.76$) and strongest Bayesian evidence ($\ln\mathcal{B} = 5.98 \pm 0.69$). Principal component analysis reveals that the data effectively constrain three additional degrees of freedom in the $\beta(z)$ model, accounting for most of the improvement in goodness of fit. Our results demonstrate that the dynamical dark energy preference in current data can be equally well explained by such a sign-reversal interacting dark energy, highlighting the need for future observations to break this degeneracy.}

\begin{document}

\maketitle
\flushbottom

\section{Introduction}\label{sec:intro}

Recent measurements of baryon acoustic oscillations (BAO) from the Dark Energy Spectroscopic Instrument (DESI)\cite{DESI:2024uvr,DESI:2024lzq}, when combined with cosmic microwave background (CMB) and type Ia supernovae (SNIa) data, suggest a hint of new physics in the nature of dark energy. The DESI Data Release 1 (DR1) results favor a time-varying equation of state (EoS) of dark energy at a significance of approximately 2.5$\sigma$ to 3.9$\sigma$, depending on the SNIa dataset used \cite{DESI:2024mwx}. This preference strengthens to 2.8$\sigma$ to 4.2$\sigma$ in the DESI DR2 analysis \cite{DESI:2025zgx}, presenting a notable challenge to the standard $\Lambda$CDM paradigm where dark energy is described by a cosmological constant $\Lambda$ with a fixed EoS of $w=-1$. The DESI data have been widely used in recent cosmological analyses; see, e.g., Refs.~\cite{Li:2024qso,Li:2025owk,Li:2024qus,Li:2025eqh,Du:2024pai,Du:2025iow,Wu:2024faw,Ling:2025lmw,Yang:2024kdo,Yang:2025mws,Li:2025cxn,Cai:2025mas,Ye:2024ywg,Wang:2024pui,Pang:2024wul,Pang:2025lvh,Pang:2024qyh,Huang:2025som,Wang:2024dka,Wang:2024hwd,Giare:2024gpk,Jiang:2024viw,Abedin:2025dis,Jiang:2024xnu,Wang:2024hks}.

While $\Lambda$CDM remains highly successful in fitting CMB data and explaining cosmic acceleration, it faces growing tensions with precision observations. A key discrepancy lies in the Hubble constant $H_0$: local distance ladder measurements yield $H_0=73.04\pm1.04$ ${\rm km}/{\rm s}/{\rm Mpc}$ \cite{Riess:2021jrx}, whereas $\Lambda$CDM-based CMB analyses infer $H_0=67.36\pm0.54$ ${\rm km}/{\rm s}/{\rm Mpc}$ \cite{Planck:2018vyg}, revealing a tension exceeding $5\sigma$. Theoretically, the cosmological constant faces long-standing fine-tuning and coincidence problems \cite{Weinberg:1988cp,Sahni:1999gb,Peebles:2002gy,Bean:2005ru,Copeland:2006wr,Sahni:2006pa,Kamionkowski:2007wv,Frieman:2008sn,Li:2011sd}. This situation motivates modifications to the $\Lambda$CDM framework, such as replacing the cosmological constant with dynamical dark energy (with an EoS not exactly equal to $-1$), or adopting modified gravity theories that avoid introducing dark energy (see Ref. \cite{Weinberg:2013agg} for a recent review). 

Cosmological models, where dark energy directly interacts with cold dark matter by exchanging energy and momentum, have attracted significant attention in recent years \cite{He:2008tn,He:2009mz,He:2009pd,Guo:2004xx,Wei:2007ut,Cai:2009ht,He:2010im,Boehmer:2008av,Guo:2007zk,Xia:2009zzb,Wei:2010cs,Li:2011ga,Zhang:2004gc,Li:2013bya,Wang:2013qy,Pourtsidou:2013nha,Yang:2015tzc,Yang:2017ccc,Cai:2017yww,Yang:2018euj,Kase:2020hst,Linton:2021cgd,Faraoni:2014vra,Zhang:2012uu,Zhang:2013lea,Valiviita:2008iv,Cai:2016hea,Abedin:2025yru,Escamilla:2023shf,Pan:2025qwy,Aboubrahim:2024cyk,DiValentino:2021izs,DiValentino:2020kpf,Silva:2025hxw,Forconi:2023hsj,Nunes:2022bhn,Wang:2016lxa} (see Ref. \cite{Wang:2024vmw} for a recent review). In such interacting dark energy (IDE) scenario, the conservation laws for the energy-momentum tensor ($T_{\mu\nu}$) of dark energy ($de$) and cold dark matter ($c$) are modified as 
\begin{equation}
  \label{eq:energyexchange} \nabla_\nu T^\nu_{\hphantom{j}\mu,de} = -\nabla_\nu T^\nu_{\hphantom{j}\mu,c}  = Q_\mu,
\end{equation} 
where $Q_\mu$ denotes the energy-momentum transfer vector. A specific $Q_{\mu}$ determines the energy transfer rate $Q$, its perturbation $\delta Q$ and the momentum transfer rate $f$. These modifications alter both the background evolution and perturbations of the dark sectors, offering a potential resolution to the coincidence problem \cite{Amendola:1999er,Comelli:2003cv,Zhang:2005rg,Cai:2004dk}, and introducing new features to the structure formation \cite{Amendola:2001rc,Bertolami:2007zm,Koyama:2009gd}.

Actually, the expansion history of the Universe in IDE and dynamical dark energy models can mimic each other by adjusting the model's parameters, which means that in this situation a dark energy model with EoS $w\neq -1$ is observationally degenerate with an IDE model with $w=-1$ but $Q\neq 0$. Given that current cosmological data show a preference for dynamical dark energy, it is natural to explore whether this deviation from $\Lambda$CDM could instead be explained by a non-zero dark sector interaction. Indeed, recent studies have explored this possibility. For instance, the authors in Ref.~\cite{Giare:2024smz} combined DESI DR1 BAO, Planck CMB, and PantheonPlus SNIa to constrain an IDE model with $Q$ proportional to the density of dark energy, finding a non-zero coupling constant exceeding the 95\% confidence level (CL). Ref. \cite{Li:2024qso} analyzed four IDE models, using similar cosmological data, reporting a non-zero coupling exceeding the 68\% CL for three cases (note that 3$\sigma$ can be reached in the case of $Q=\beta H_0\rho_{de}$). In our recent work \cite{Li:2025owk}, we further examined whether current data support a sign-changed interaction. 

While these studies provide valuable insights, their conclusions remain inherently model-dependent. In this work, we present a model-independent reconstruction of the vacuum energy ($w=-1$) interacting with cold dark matter, following the non-parametric framework of Refs. \cite{Crittenden:2005wj,Zhao:2012aw}. By discretizing the coupling function and assuming a Gaussian smoothness prior, we reconstruct the coupling without assuming any specific parameterization. We utilize the most recent cosmological observations, including DESI DR1/DR2 BAO measurements, Planck CMB data, and three distinct SNIa samples. This comprehensive analysis enables us to probe the possible dark sector interaction without theoretical bias. We will show that the dynamical dark energy preference in current data can be equally well explained by IDE scenario. 
 
The structure of this paper is organized as follows. In Section~\ref{sec:GeneralEqs}, we briefly introduce the IDE model. In Section~\ref{sec:method}, we outline how we implement the non-parametric reconstruction. In Section~\ref{sec:Data}, we show the cosmological data used in this work, and report the reconstruction results. The conclusion is given in Section~\ref{sec:conclusions}.

\section{The model}\label{sec:GeneralEqs}
The conservation laws in the background level for the densities of dark energy ($\rho_{de}$) and cold dark matter ($\rho_{c}$) satisfy
\begin{gather}
  \rho'_{de} +3\mathcal{H}(1+w)\rho_{de}= aQ,\label{eq:rhodedot}\\
  \qquad\rho'_{c} +3\mathcal{H}\rho_{c}=- aQ,\label{eq:rhocdot}
\end{gather}
where $a$ is the scale factor of the Universe, $w={p_{de}/\rho_{de}}$ is the EoS of dark energy, $\mathcal{H}=a'/a$ is the conformal Hubble parameter, and a prime denotes the derivative with respect to the conformal time. In this paper, we set the EoS of dark energy $w=-1$. Namely, we consider the case of vacuum energy interacting with cold dark matter. We assume the energy transfer rate $Q$ to be of the following form
\begin{equation}
  Q = \beta(a)H\rho_{de},\label{eq:Qenergy}
\end{equation}
where $\beta(a)$ is the dimensionless coupling parameter (that is evolutionary) and $H$ is the Hubble parameter. The specific form of $Q$ is chosen for mathematical convenience without loss of generality, as any arbitrariness in this choice is fully absorbed into the time-dependent coupling function $\beta(a)$. We have verified that alternative forms (e.g., $Q\propto\rho_{c}$) yield equivalent reconstruction results. For clarity in subsequent discussion, we refer to this framework as the $\beta(a)$ model (or equivalently, the $\beta(z)$ model when expressed in terms of cosmological redshift).

In this study, we avoid imposing any ad hoc parameterization on the coupling function $\beta(a)$. Instead, we reconstruct it from cosmological data using non-parametric approach. Specifically, we discretize the evolution of $\beta$ by dividing the scale factor interval $[a_{\rm min},1]$ into $N$ bins, treating $\beta$ as piecewise constant within each bin, i.e., 
\begin{equation}
  \beta(a)=\sum_{i=1}^N \beta_i T_i(a),
\quad 
T_i(a)=\begin{cases}1,&a_{i}\leq a<a_{i+1},\\0,&\text{otherwise},\end{cases}
\end{equation}
where $\beta_i$ for $i=1,2,...,N$ denotes the amplitude of each bin. {We set $a_1=a_{\rm min}=0.0001$ (corresponding to $z \sim 10^4$), $a_{N+1}=1$ and $\beta(a)=0$ at $a<a_1$. This choice, which serves as an initial condition at a very early epoch, implies no interaction between dark energy and dark matter in the primordial Universe. It is a physically reasonable assumption, as the energy densities of both dark energy and dark matter are subdominant in the extreme early Universe, making any interaction between them negligible.} Then the model's background can be solved, and the energy densities of the dark sectors evolve as
\begin{equation}
  \rho_{de}=\rho_{de0}\prod_{k=i+1}^{N}{a_k^{\beta_k -\beta_{k-1}}}a^{\beta_{i}},
\end{equation} 
and
\begin{align}
  \rho_{c}&= \rho_{de0}a^{-3}\sum_{j=i}^{N-1}\prod_{k=j+1}^{N}{a_k^{\beta_k -\beta_{k-1}}} \frac{\beta_{j}}{\beta_{j} + 3}\left(a_{j+1}^{\beta_{j}+3}-a_{j}^{\beta_{j}+3}\right) \nonumber \\ 
    &+\rho_{de0}a^{-3}\prod_{k=i+1}^{N}{a_k^{\beta_k -\beta_{k-1}}} \frac{\beta_{i}}{\beta_{i} + 3}\left(a_{i}^{\beta_{i}+3}-a^{\beta_{i}+3}\right) \nonumber \\
    &+\rho_{c0}a^{-3} +\rho_{de0}a^{-3}\frac{\beta_{N}}{\beta_{N} + 3}(1-a_N^{\beta_N+3}),
\end{align}
at $a_i\leq a<a_{i+1}$ for $i=1,2,...,N$. Here we set $\sum_{j=N}^{N-1}=0$ and $\prod_{k=N+1}^{N}=1$ for $i=N$ ($a_N\leq a<1$).

For the model's perturbation, we assume the energy-momentum transfer $Q_\mu$ to be parallel to the four-velocity of dark matter, i.e., $Q_\mu=Qu_{\mu,c}$. In the choice, the momentum transfer rate $f$ vanishes in the dark matter rest frame, and hence the dark matter particles follow geodesics. For dark energy, we handle its perturbation using the extended parameterized post-Friedmann (ePPF) approach \cite{Li:2014eha}, which is the PPF framework \cite{Hu:2008zd,Fang:2008sn} applied to IDE theory. The ePPF approach can help avoid the possible large-scale instability \cite{Valiviita:2008iv,He:2008si,Clemson:2011an} that may occur in standard linear perturbation theory at specific parameter values. In previous works \cite{Li:2014eha,Li:2014cee}, we established the ePPF approach for different IDE models. More recently, we further developed an ePPF approach for a general parametrization IDE model with five free functions $C_1$, $C_2$, $C_3$, $D_1$ and $D_2$ \cite{Li:2023fdk}. This model-independent formulation allows different IDE scenarios to be mapped onto the parametrization by specifying these functions. For our specific $\beta(a)$ model, the corresponding forms of the five functions are $C_1=D_2=Q$ and $C_2=C_3=D_1=0$. Further details on the parametrization and the ePPF implementation can be found in Ref.~\cite{Li:2023fdk}.

\section{The reconstruction method}\label{sec:method}
To reconstruct the coupling history, $\beta(a)$, we can constrain the amplitude of each bin, $\beta_i$, using the data. However, if the neighboring bins are treated as independent, fitting a large number of uncorrelated bins could result in excessively large uncertainties. To avoid this issue, we adopt the non-parametric approach developed in Refs. \cite{Crittenden:2005wj,Zhao:2012aw}, in which the variable to be reconstructed is assumed to be a smooth function, and hence its values at neighboring points are not entirely independent. This feature for $\beta(a)$ can be described by a correlation function between its values at $a$ and $a^\prime$,
\begin{equation}
\xi(|a-a^\prime|) \equiv \left<  [\beta(a)-\beta^{\rm fid}(a)][\beta(a^\prime)-
\beta^{\rm fid}(a^\prime)] \right>,
\end{equation}
with $\beta^{\rm fid}$ a reference fiducial model. Here $\beta(a)$ is assumed to be a Gaussian random variable with a covariance matrix among a large number of bins given by
\begin{equation}\label{eq:Cij} 
  C_{ij}=\frac{1}{\Delta^2}\int_{a_i}^{a_i+\Delta}da \int_{a_j}^{a_j+\Delta}da'~\xi (|a - a'|),
\end{equation}
where $\Delta$ is the bin width. This covariance matrix defines a Gaussian prior probability distribution for $\beta_i$, i.e.,
\begin{equation}\label{eq:prior}
\mathcal{P}_{\rm prior}(\bm{ \beta}) \propto \exp\left( -\frac{1}{2}(\bm{\beta}-\bm{\beta}^{\rm fid})
\bf{C}^{-1} (\bm{\beta}-\bm{\beta}^{\rm fid} ) \right).
\end{equation}
According to Bayes' theorem, the desired posterior probability is proportional to the likelihood of the data times the prior probability, 
\begin{equation}
{\cal P}(\bm{\beta}|{\bf D}) \propto {\cal P}({\bf D}|\bm {\beta}) \times {\cal P}_{\rm prior}(\bm {\beta}),
\end{equation}
which penalizes those models that are less smooth. 

The calculation of the covariance matrix depends on a specific $\xi(|a-a^\prime|)$. Following Ref.~\cite{Crittenden:2011aa}, we assume the correlation function to have the form proposed by Crittenden, Pogosian and Zhao (CPZ) \cite{Crittenden:2005wj}, 
\begin{equation}\label{eq:CPZ}
\xi_{\rm CPZ}(\delta a) =  \frac{\xi (0)} {1 + (\delta a/a_c)^2},
\end{equation}
where $\delta a\equiv|a-a^\prime|$, $a_c$ describes the typical smoothing scale, and $\xi(0)$ is the normalization factor determined by the expected variance of the mean of $\beta$, i.e., $\sigma^2_{\beta} \simeq \pi\xi(0){a_c}/(1-a_{\rm min})$. One can adjust the prior strength by tuning the values of $a_c$ and $\sigma^2_{\beta}$. A weak prior (i.e., small $a_c$ or large $\sigma^2_{\beta}$) can match the true model on average, but will result in a noisy reconstruction, while a stronger prior reduces the variance but pulls the reconstructed results towards the peak of the prior. In this paper, we set $a_c=0.06$ and $\sigma_{\beta}=0.04$, representing a moderate prior, which was also adopted in Ref.~\cite{Wang:2015wga}. To avoid the explicit dependence on the fiducial model, one can marginalize over the value of $\bm{\beta}^{\rm fid}$ with several ways \cite{Crittenden:2011aa}. In this paper, we use the `floating' average method \cite{Crittenden:2011aa}, which takes the fiducial model to be a local average,
\begin{equation}
\beta^{\rm fid}_i = \sum_{|a_j - a_i| \leq a_c} \beta^{\rm true}_j / N_j,
\end{equation}
where $N_j$ is the number of neighboring bins around the $i$-th bin within the smoothing scale. 

{To validate our reconstruction framework and ensure that the inferred features of $\beta(z)$ are driven by the data rather than artefacts of the non-parametric method, we performed a test using a noiseless synthetic dataset. We generated simulated BAO and SNIa data based on the $\Lambda$CDM cosmology (i.e., with $\beta(z)=0$). Applying our full reconstruction pipeline to this synthetic dataset, we recovered a $\beta(z)$ function consistent with zero within $1\sigma$ across all redshifts, as expected. The details of this validation test are presented in Appendix~\ref{app:validation}.}

\section{Data and results}\label{sec:Data}
In this work, we constrain the amplitudes of 20 bins, i.e., $N=20$, with the first bin covering the wide range of $a\in[0.0001,0.286]$, and the last 19 bins uniform in $a\in[0.286,1]$, corresponding to $z\in[0,2.5]$. {We have verified that our central conclusions are robust against increasing the number of bins (e.g., $N=30$), or extending the redshift range of the uniform bins to higher redshifts (e.g., from $z=2.5$ to $z=4.5$). The results of these tests are detailed in Appendix~\ref{app:robustness}.} We sample the parameter space using the public package {\tt Cobaya} \cite{Torrado:2020dgo} with the nested sampler {\tt PolyChord} \cite{Handley:2015fda,Handley:2015vkr}. The obtained samples are analyzed using the public package {\tt Getdist} \cite{Lewis:2019xzd}. The Gaussian prior defined in Eq.~(\ref{eq:prior}) is computed as an external prior in {\tt Cobaya}. The theoretical model is solved using the {\tt IDECAMB} code \cite{Li:2023fdk}, which is an implementation of IDE theory in the public Einstein-Boltzmann solver {\tt CAMB} \cite{Lewis:1999bs,Howlett:2012mh}. To ensure compatibility with the {\tt Cobaya} package, we have rewritten the {\tt IDECAMB} code based on the new version of {\tt CAMB}. To ensure numerical stability in nested sampling, we adopt conservative priors on the cosmological parameters: the physical baryon density $\Omega_bh^2 \in [0.01, 0.05]$, the physical cold dark matter density $\Omega_ch^2 \in [0.02, 0.3]$, the Hubble constant $H_0 \in [40, 100]$, the reionization optical depth $\tau_\mathrm{reio} \in [0.01, 0.05]$, the amplitude and spectral index of the primordial scalar spectrum $\mathrm{ln}(10^{10}A_s) \in [2.5, 3.5]$ and $n_s \in [0.85, 1.1]$, and the coupling constants $\beta_i \in [-2, 2]$ for $i=1,2,...,20$. We take $H_0$ as a free parameter instead of the commonly used $\theta_{\rm{MC}}$, because $\theta_{\rm{MC}}$ is dependent on a standard non-interacting background evolution.

We adopt the following datasets:

$\bullet$ CMB: For the CMB data, we use the temperature (TT), polarization (EE), cross (TE), and lensing power spectra measurements. The specific likelihoods are: (i) the \texttt{Commander} likelihood using the temperature spectrum at large scales ($2 \leq \ell \leq 30$), from the Planck 2018~\cite{Planck:2018vyg,Planck:2019nip}; (ii) the \texttt{SimAll} likelihood using the spectrum of E-mode polarization at large scales ($2 \leq \ell \leq 30$) from the Planck 2018~\cite{Planck:2018vyg,Planck:2019nip}; (iii) the high-$\ell$ \texttt{CamSpec} ($\ell \geq  30$) likelihood using dust-cleaned TT, TE, and EE power spectra computed from the Planck 2020 (PR4 / NPIPE) maps \cite{Rosenberg:2022sdy}; (iv) the CMB lensing likelihood from the NPIPE PR4 Planck \cite{Carron:2022eyg}. We label the combination of these likelihoods as ``CMB". 

$\bullet$ BAO: We utilize the BAO measurements from the DESI DR2 \cite{DESI:2025zgx}, which includes tracers from the bright galaxy sample (BGS), luminous red galaxies (LRG), emission line galaxies (ELG), quasars (QSO), and the Ly$\alpha$ forest in a redshift range $0.1\leq z \leq 4.2$. These tracers are described through the transverse comoving distance $D_{\mathrm{M}}/r_{\mathrm{d}}$, the angle-averaged distance $D_{\mathrm{V}}/r_{\mathrm{d}}$, and the Hubble horizon $D_{\mathrm{H}}/r_{\mathrm{d}}$, where $r_{\mathrm{d}}$ is the comoving sound horizon at the drag epoch. We also compare with the results from DR1 \cite{DESI:2024mwx}.


$\bullet$ SNIa: We adopt SNIa data from three compilations: (i) the PantheonPlus (PP) dataset, which comprises 1550 spectroscopically confirmed SNIa in the redshift range of $0.001 < z < 2.26$ \cite{Brout:2022vxf} (The public likelihood imposes a bound of $z > 0.01$ in order to mitigate the impact of peculiar velocities in the Hubble diagram); (ii) the Dark Energy Survey Year 5 (DESY5) data release, which includes 194 low-redshift SNIa ($0.025 < z < 0.1$) and 1635 SNIa classified photometrically, covering the range $0.1 < z < 1.3$ \cite{DES:2024tys}; (iii) the more recent Union3 compilation, presented in Ref.~\cite{Rubin:2023ovl}, including 2087 SNIa, with many (1363 SNIa) overlapping with those in PantheonPlus. These SNIa datasets are not independent of each other. Therefore, we do not combine them but instead present results from CMB+DESI plus each SNIa dataset independently.

\begin{figure*}[!htbp]
  \includegraphics[width=0.95\textwidth]{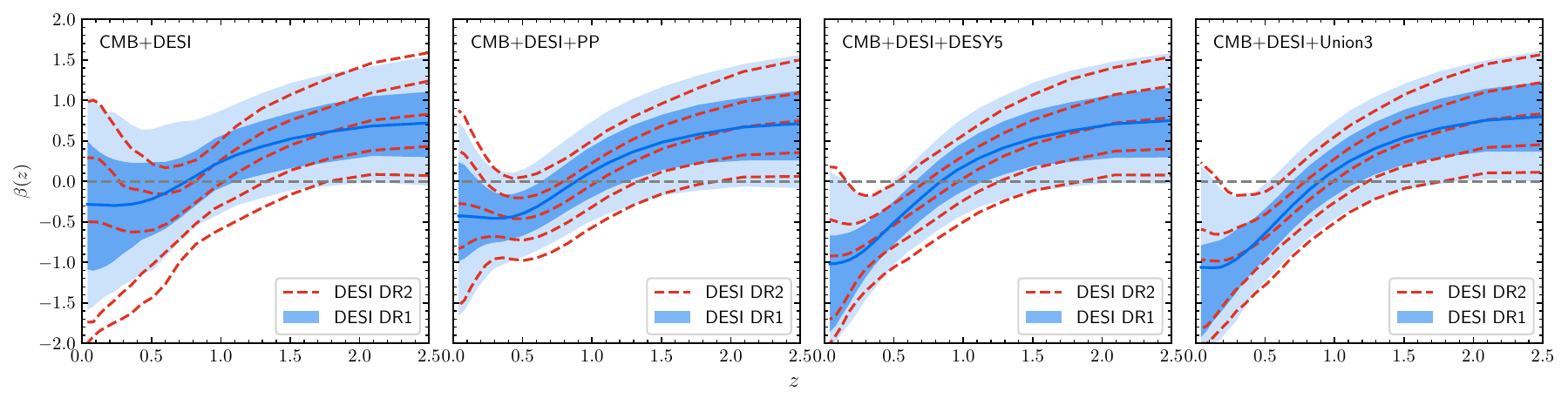}
  \centering
  \caption{\label{fig:betaz} Reconstructed evolution history of $\beta(z)$ with the mean value and 68\% and 95\% CL errors for the baseline CMB+DESI and its individual combinations with each SNIa dataset (PP, DESY5, and Union3). The blue solid lines and filled regions represent the DESI DR1-based reconstructions, while the red dashed lines show the DR2-based results. The black dashed lines denote the $\Lambda$CDM prediction ($\beta=0$).}
  \end{figure*}

Convergence in PolyChord is achieved when the fraction of the total evidence contained in the live points  falls below 0.001. In Fig.~\ref{fig:betaz}, we plot the reconstructed $\beta(z)$ with the mean value and 68\% and 95\% CL errors for the baseline CMB+DESI and its individual combinations with each SNIa dataset (PP, DESY5, and Union3). The blue solid lines and filled regions represent the DESI DR1-based reconstructions, while the red dashed lines show the DR2-based results. A key feature across all reconstructions is that the mean value of $\beta(z)$ changes sign during its evolution. Specifically, it is positive at early times, suggesting an energy transfer from cold dark matter to dark energy, but turns negative at late times, implying an energy flow in the reverse direction. This behavior is consistent with the findings of Ref.~\cite{Li:2025owk}, in which the coupling is reconstructed with a parametric model. 

When reconstruction uncertainties are taken into account, all data combinations yield a positive $\beta(z)$ that deviates from the $\Lambda$CDM prediction ($\beta=0$) at approximately $2\sigma$ at high redshifts. At low redshifts, however, the results vary depending on whether and which SNIa dataset is used. The CMB+DESI and the CMB+DESI+PP data get a $\beta(z)$ consistent with a zero coupling within $2\sigma$, whereas CMB+DESI+DESY5 and CMB+DESI+Union3 show a preference for negative $\beta$ at about $2\sigma$ level. The DESI DR2-based results are consistent with those from DR1, but show lower reconstruction uncertainties in the redshift range $0.3 < z < 2.5$. This improvement aligns with the primary redshift coverage of DESI observations and reflects the enhanced precision achieved in DR2.

\begin{table*}[!htbp]
  \renewcommand\arraystretch{1.5}
  \centering
  \caption{Marginalized mean values and 68\% CL intervals for the parameters of interest.}\label{tab:params}
\begin{tabular} { l  c c c c}
  \hline
  Parameter &  CMB+DESI DR2 &  +PP  &  +DESY5 &  +Union3\\
 \hline
 $H_0            $ & $67.2^{+1.2}_{-1.8}        $ & $67.59\pm 0.58             $ & $66.84\pm 0.55             $ & $66.45^{+0.68}_{-0.84}     $\\
 
 $\Omega_{m}         $ & $0.393^{+0.150}_{-0.095}    $ & $0.363\pm 0.049            $ & $0.437^{+0.058}_{-0.032}   $ & $0.456^{+0.073}_{-0.030}   $   \\

 $\sigma_8                  $ & $0.716^{+0.030}_{-0.220}    $ & $0.706^{+0.063}_{-0.110}    $ & $0.602^{+0.028}_{-0.075}   $ & $0.587^{+0.016}_{-0.082}   $\\
 
 $S_8$ & $0.770^{+0.022}_{-0.100}$ & $0.769^{+0.037}_{-0.051}$ & $0.721^{+0.018}_{-0.042}$ & $0.715^{+0.012}_{-0.04}$\\
 \hline
 \end{tabular}
 
\end{table*}

{The reconstruction of the coupling function $\beta(z)$ directly impacts the evolution of the Universe's expansion history and matter clustering. It is therefore crucial to examine the constraints on other key cosmological parameters derived within this interacting scenario. In Table~\ref{tab:params}, we present the marginalized mean values and 68\% CL intervals for the Hubble constant $H_0$, the matter density parameter $\Omega_m$, the amplitude of matter fluctuations $\sigma_8$ and $S_8\equiv \sigma_8(\Omega_m/0.3)^{0.5}$, focusing on the data combinations with DESI DR2. The derived $H_0$ values, ranging from $66.45$ to $67.59$ $\mathrm{km/s/Mpc}$, show no significant deviation from the Planck $\Lambda$CDM expectation, indicating that the Hubble tension is not alleviated in the $\beta(z)$ model. In contrast, the $S_8$ values, ranging from $0.715$ to $0.770$, are systematically lower than the typical Planck $\Lambda$CDM value ($S_8 \approx 0.832$). This suppression of structure growth represents a moderate alleviation of the $S_8$ tension.}

To assess the improvements in goodness of fit, we compute $\Delta\chi_{\rm MAP}^2$, the difference of $\chi^2$ in the maximum a posteriori (MAP) between the $\beta(z)$ model and the $\Lambda$CDM model, using the minimizer sampler \cite{BOBYQA,Cartis:2018xum,Cartis:2018jxl} in {\tt Cobaya}. In Table \ref{tab:results}, we summarize the $\Delta\chi_{\rm MAP}^2$ values for the different data combinations. Across all dataset combinations, the $\beta(z)$ model yields a better fit than $\Lambda$CDM, with the most notable improvement occurring for CMB+DESI DR2+DESY5. For the baseline $\mathrm{CMB}+\mathrm{DESI}$ analysis, we observe $\Delta\chi_{\rm MAP}^2 = -5.46$ (DR1) and $-4.37$ (DR2). The inclusion of SNIa data consistently enhances the preference for the $\beta(z)$ model over $\Lambda$CDM, though the degree of improvement varies across datasets. For the $\mathrm{CMB}+\mathrm{DESI}+\mathrm{PP}$ data combination, we find moderate improvements in fit quality, with $\Delta\chi_{\rm MAP}^2 = -5.49$ for DR1 and $-7.88$ for DR2. In contrast, DESY5 demonstrates the most pronounced preference, yielding $\Delta\chi_{\rm MAP}^2 = -13.45$ (DR1) and $-17.76$ (DR2). Similarly, Union3 shows strong but slightly weaker improvements, with $\Delta\chi_{\rm MAP}^2 = -10.94$ (DR1) and $-9.78$ (DR2).

  \begin{table}[!htbp]
    \renewcommand\arraystretch{1.5}
    \centering
    \caption{Summary of the $\Delta\chi_{\rm MAP}^2$ values, the difference of $\chi^2$ in the maximum a posteriori (MAP) between the $\beta(z)$ model and the $\Lambda$CDM model, for the different combinations of datasets as indicated. The third column lists the corresponding Bayes factor in logarithmic space, $\ln \mathcal{B} = \ln Z_{\beta(z)} - \ln Z_{\rm \Lambda CDM}$, where $\ln Z_{\beta(z)}$ and $\ln Z_{\rm \Lambda CDM}$ represent the Bayesian evidence of the $\beta(z)$ model and $\Lambda$CDM, respectively.}
    \label{tab:results}
    \begin{tabular} { l  c c c c }
      \hline
      Datasets &  $\Delta\chi_{\rm MAP}^2$ &  $\ln \mathcal{B}$  \\
     \hline
     CMB+DESI DR1 & $-5.46$ & $1.36\pm 0.68             $ \\
  
     CMB+DESI DR2 & $-4.37$ & $1.44\pm 0.70            $ \\
  
     CMB+DESI DR1+PP & $-5.49$ & $-0.19\pm 0.64             $ \\
  
     CMB+DESI DR2+PP & $-7.88$ & $1.12\pm 0.69            $ \\
  
     CMB+DESI DR1+DESY5 & $-13.45$ & $3.42\pm 0.68             $\\
  
     CMB+DESI DR2+DESY5 & $-17.76$ & $5.98\pm 0.69            $ \\
  
     CMB+DESI DR1+Union3 & $-10.94$ & $1.92\pm 0.70             $ \\
  
     CMB+DESI DR2+Union3 & $-9.78$ & $3.57\pm 0.69            $\\
     
     \hline
     \end{tabular}
    \end{table}

    \begin{figure}[!htbp]
      \includegraphics[width=0.7\textwidth]{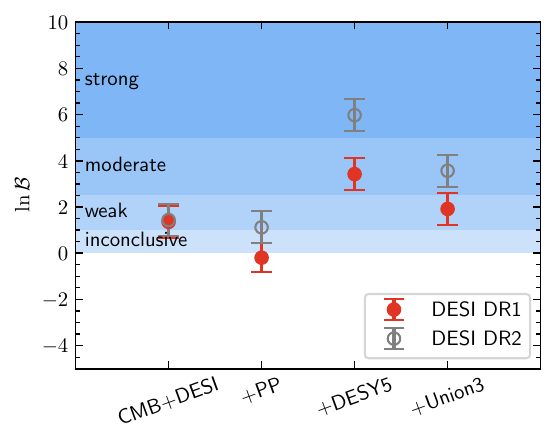}
      \centering
      \caption{\label{fig:bayes} The Bayes factor $\ln \mathcal{B}$ results with 68\% CL errors for different data combinations. The shaded regions correspond to the Jeffreys' scale thresholds for evidence interpretation: $|\ln\mathcal{B}|<1$ (inconclusive), $1\leq|\ln\mathcal{B}|<2.5$ (weak), $2.5\leq|\ln\mathcal{B}|<5$ (moderate), and $|\ln\mathcal{B}|\geq5$ (strong).}
      \end{figure}

However, the $\chi^2$ test does not account for model complexity or the number of parameters involved. To address this limitation, we further compute the Bayes factor in logarithmic space, $\ln \mathcal{B} = \ln Z_{\beta(z)} - \ln Z_{\rm \Lambda CDM}$, where $\ln Z_{\beta(z)}$ and $\ln Z_{\rm \Lambda CDM}$ represent the Bayesian evidence of the $\beta(z)$ model and $\Lambda$CDM, respectively. The Bayesian evidence is directly obtained from the PolyChord sampler output, though we note that the external prior defined in Eq.~(\ref{eq:prior}) is unnormalized by default. Thus, the raw evidence $Z_{u}$ must be corrected by the prior volume $Z_{\pi}$ as $\ln Z_{\beta(z)}=\ln Z_{u} - \ln Z_{\pi}$. The $\ln \mathcal{B}$ results with 68\% CL errors are summarized in Table \ref{tab:results} and shown in Fig.~\ref{fig:bayes}. For visual reference, Fig.~\ref{fig:bayes} includes shaded regions corresponding to the Jeffreys' scale thresholds \cite{Kass:1995loi,Trotta:2008qt} for evidence interpretation. Our analysis demonstrates a consistent trend of stronger evidence for the $\beta(z)$ model in DESI DR2 compared to DR1 across all data combinations. For $\mathrm{CMB}+\mathrm{DESI}$ $\mathrm{DR1}/\mathrm{DR2}$, we find $\ln\mathcal{B} = 1.36\pm0.68/1.44\pm0.70$, indicating \textit{weak evidence} in both cases, with DR2 showing slightly stronger (but statistically consistent) preference. The inclusion of PP SNIa yields $\ln\mathcal{B} = -0.19\pm0.64$ (DR1) and $1.12\pm0.69$ (DR2) -- while DR1 remains \textit{inconclusive}, DR2 crosses the threshold into \textit{weak evidence} territory. Most significantly, $\mathrm{CMB}+\mathrm{DESI}+\mathrm{DESY5}$ shows $\ln\mathcal{B} = 3.42\pm0.68$ (DR1) and $5.98\pm0.69$ (DR2). Here, DR1 provides \textit{moderate evidence} while DR2 reaches \textit{strong evidence}. {Similarly, $\mathrm{CMB}+\mathrm{DESI}+\mathrm{Union3}$ transitions from \textit{weak} ($1.92\pm0.70$) to \textit{moderate} ($3.57\pm0.69$) evidence from DR1 to DR2.}

The positive values of $\ln \mathcal{B}$ imply that the improved goodness of fit in the $\beta(z)$ model outweighs the Occam's razor penalty arising from parameter space integration. This implies that only a subset of the additional parameters is effectively constrained by the data. To quantify this, we assess the number of effective degrees of freedom introduced by the $\beta(z)$ model relative to $\Lambda$CDM. We perform a principal component analysis (PCA) on the posterior distribution by diagonalizing the covariance matrix of the $\beta$ bins after marginalizing over other cosmological parameters. The eigenvectors $e_i(z)$ form a basis for expanding $\beta(z)$ as $\beta(z) = \sum_{i=1}^N \alpha_i e_i(z)$, while the eigenvalues $\lambda_i$ determine the measurement precision of each mode, with $\sigma(\alpha_i) = \sqrt{\lambda_i}$ \cite{Huterer:2002hy}. 

\begin{figure}[!htbp]
  \includegraphics[width=0.7\textwidth]{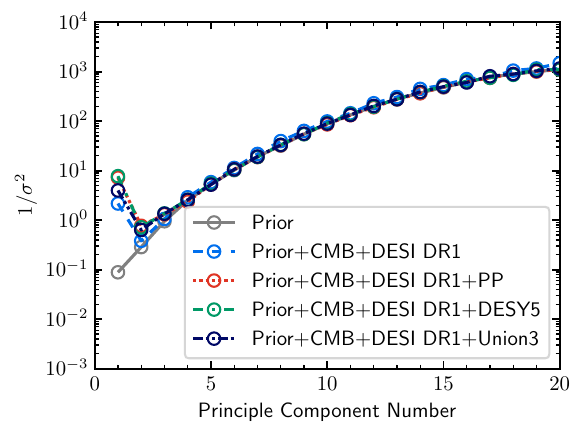}
  \centering
  \caption{\label{fig:evals} The inverse eigenvalues of both the prior and the four posterior covariance matrices, ordered by the number of nodes in $e_i(z)$.}
  \end{figure}

\begin{figure}[!htbp]
    \includegraphics[width=0.7\textwidth]{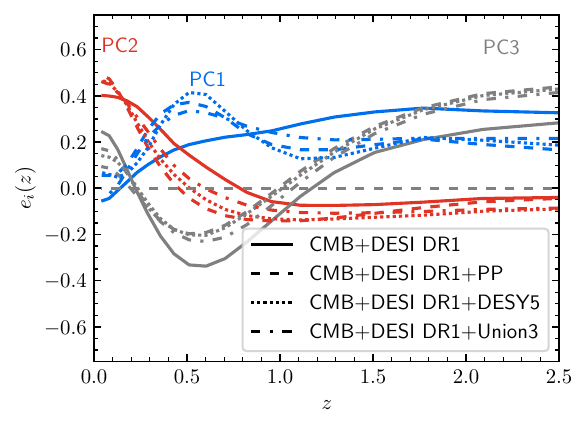}
    \centering
    \caption{\label{fig:pc} The first three data-dominated eigenvectors $e_i(z)$ for the four DESI DR1-based data combinations.}
    \end{figure}

As a representative example, Fig.~\ref{fig:evals} displays the inverse eigenvalues of the posterior covariance matrices for the four DESI DR1-based data combinations, ordered by the number of nodes in the eigenvectors $e_i(z)$. For comparison, we also plot the inverse eigenvalues derived from the prior-only covariance matrix. The results reveal a clear distinction: while the first three posterior inverse eigenvalues significantly exceed their prior counterparts, the fourth and higher eigenvalues show close agreement with the prior values. This demonstrates that only the first three principal components are data-constrained, whereas the higher-order modes remain prior-dominated. We therefore conclude that the $\beta(z)$ model effectively introduces three additional degrees of freedom compared to $\Lambda$CDM. These three data-dominated eigenvectors $e_i(z)$ are presented in Fig.~\ref{fig:pc}. By projecting the mean $\beta(z)$ onto these three data-dominated eigenvectors, i.e., $\beta(z)=\sum_{i=1}^{3}\alpha_ie_i(z)$, we derive the constraints on the coefficients $\alpha_i$ (mean values and 68\% CL errors), as summarized in Table \ref{tab2}. The improvement in fit $\Delta\chi^2_{\rm MAP}$ is well approximated by $\sum_{i=1}^3 [\alpha_i / \sigma(\alpha_i)]^2$, confirming that these three principal components account for nearly all the enhanced model performance.          

  \begin{table}[!htbp]

    \renewcommand\arraystretch{1.5}
    \centering
    \caption{The coefficients $\alpha_i$ (mean values and 68\% CL errors) of the first three data-dominated modes for the four DESI DR1-based data combinations.}
    \label{tab2}
    \begin{tabular} { c|  c| c| c |c }
      \hline
                 &  CMB+DESI &  +PP  &  +DESY5 &  +Union3\\
     \hline
     
                &$-1.16 \pm 0.68$ & $-0.35 \pm 0.37$ & $-0.65 \pm 0.36$ & $0.44 \pm 0.50$ \\
     $\alpha_i$ &$-1.05 \pm 1.62$ & $-1.08 \pm 1.16$ & $-2.67 \pm 1.18$ & $-2.99 \pm 1.25$ \\
                &$0.92 \pm 0.97$ & $1.80 \pm 0.87$ & $1.42 \pm 0.85$ & $1.53 \pm 0.86$ \\
          
     \hline
     \end{tabular}
    \end{table}

\section{Conclusion}\label{sec:conclusions}

In this study, we conduct a model-independent reconstruction of the vacuum energy ($w=-1$) interacting with cold dark matter, using the latest cosmological data, including DESI DR1/DR2 BAO measurements, Planck CMB data, and three distinct SNIa datasets (PP, DESY5, and Union3). By employing a non-parametric approach with a Gaussian smoothness prior, we reconstruct the coupling function $\beta(z)$ without assuming any specific parameterization. The results consistently show a sign change in the mean value of $\beta(z)$ during cosmic evolution, transitioning from positive values at early times (indicating energy transfer from dark matter to dark energy) to negative values at late times (suggesting a reversed energy flow). When considering reconstruction uncertainties, all data combinations yield a positive $\beta$ that deviates from the $\Lambda$CDM prediction ($\beta=0$) at approximately $2\sigma$ at high redshifts. At low redshifts, the CMB+DESI and the CMB+DESI+PP data get a $\beta$ consistent with a zero coupling within $2\sigma$, whereas CMB+DESI+DESY5 and CMB+DESI+Union3 show a preference for negative $\beta$ at about $2\sigma$ level. The DESI DR2-based results are consistent with those from DR1, but show slightly reduced uncertainties. The $\beta(z)$ model provides a better fit to the data compared to $\Lambda$CDM, as evidenced by improvements in $\Delta\chi^2_{\mathrm{MAP}}$ and Bayesian evidence ($\ln \mathcal{B}$). Notably, the CMB+DESI DR2+DESY5 data yield the strongest evidence for the sign-reversal IDE scenario, reaching the ``strong'' threshold on the Jeffreys' scale. Principal component analysis reveals that the data effectively constrain three additional degrees of freedom in the $\beta(z)$ model, accounting for most of the improvement in goodness of fit. 

While current observations show intriguing hints of dynamical dark energy, our analysis demonstrates that these signatures could equally originate from dark sector interactions. {For the most deviating data combination (CMB+DESI DR2+DESY5), both the CPL model ($\Delta\chi^2_{\rm MAP} = -19.43$, $\ln \mathcal{B} = 6.17 \pm 0.61$) and our $\beta(z)$ model ($\Delta\chi^2_{\rm MAP} = -17.76$, $\ln \mathcal{B} = 5.98 \pm 0.69$) provide significant improvements over $\Lambda$CDM, with the CPL model offering a marginally better fit to the data. This degeneracy in background evolution presents a fundamental challenge in understanding the true nature of cosmic acceleration, as the expansion histories from both models can mimic each other through parameter adjustments. However, the distinct nature of these models becomes apparent in their predictions for structure formation. This is clearly reflected in the $S_8$ values from the CMB+DESI DR2+DESY5 combination: our reconstructed IDE model predicts a substantially lower $S_8$ value of $\sim 0.721$, while the corresponding CPL constraint gives $S_8 = 0.8267 \pm 0.0086$, much closer to the Planck $\Lambda$CDM value. Although $S_8$ provides an integrated measure of matter clustering at a specific scale, this stark contrast indicates fundamentally different structure growth histories between the two models. Future observations that probe the growth history across wider ranges of scales and redshifts (e.g., with weak lensing and redshift-space distortions) will therefore be crucial to distinguish between these two compelling explanations for the observed cosmic acceleration.}

\begin{acknowledgments}
  We thank Tian-Nuo Li and Guo-Hong Du for their helpful discussions. This work was supported by the National SKA Program of China (Grants Nos. 2022SKA0110200 and 2022SKA0110203), the National Natural Science Foundation of China (Grants Nos. 12473001, 11975072, 11835009, and 11805031), and the China Manned Space Program (Grant No. CMS-CSST-2025-A02).
\end{acknowledgments}

\appendix
\section{Validation with Synthetic Data} \label{app:validation}

\begin{figure}[!htbp]
    \centering
    \includegraphics[width=0.7\textwidth]{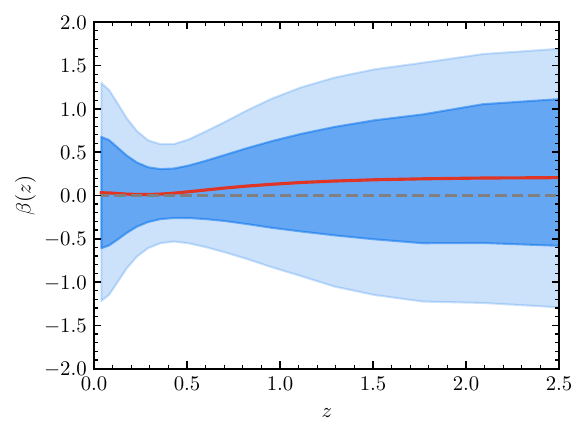}
    \caption{Reconstruction of $\beta(z)$ with the mean value (red line) and 68\% and 95\% CL errors (blue band) from a noiseless synthetic dataset generated assuming a $\Lambda$CDM cosmology ($\beta(z)=0$). The black dashed line indicates the true underlying model ($\beta=0$). The reconstruction is fully consistent with the null hypothesis, demonstrating the absence of significant biases or artefacts in our method.}
    \label{fig:validation}
\end{figure}

{Non-parametric approaches are very powerful, but the lack of explicit modelling may introduce degeneracies and reconstruction artefacts. To address this concern, we perform a validation test using a noiseless synthetic dataset. The importance of validating such reconstructions with synthetic data has been highlighted in the literature (see, e.g., Appendix E of Ref.~\cite{DESI:2025fii}). This appendix presents the test and its results, demonstrating the robustness of our reconstruction pipeline.

We generated simulated BAO and SNIa data by assuming a fiducial $\Lambda$CDM cosmology (Planck 2018 parameters \cite{Planck:2018vyg}), where by definition the interaction is absent ($\beta^{\rm fid}(z) \equiv 0$). We then applied our full reconstruction methodology, as described in Section~\ref{sec:method}, to this ideal dataset. Figure~\ref{fig:validation} shows the reconstruction result. The reconstructed $\beta(z)$ is consistent with zero across the entire redshift range, with the 68\% and 95\% CL encompassing the true model. This successful recovery of the null signal validates our reconstruction framework. }

\section{Robustness of the Reconstruction} \label{app:robustness}

{This appendix presents robustness tests for our non-parametric reconstruction. Inspired by the validation of redshift range sensitivity in Ref. \cite{DESI:2024aqx}, we systematically examine the impact of both the reconstruction range and the number of redshift bins, confirming our results are robust to these analysis choices.}

\begin{figure*}[!htbp]
  \centering
  \includegraphics[width=0.45\textwidth]{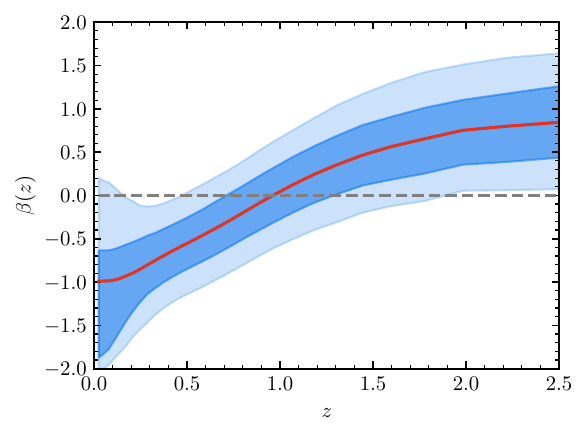}
  \includegraphics[width=0.45\textwidth]{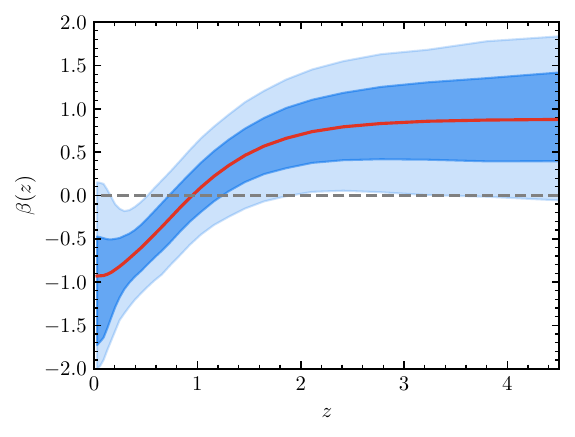}
  \caption{Robustness tests of the $\beta(z)$ reconstruction using the CMB+DESI DR2+DESY5 dataset. (left) Reconstruction with 30 bins, showing the same sign-reversal pattern as the 20-bin baseline. (right) Reconstruction with uniform binning extended to $z=4.5$, maintaining consistency at $z<2.5$ with the baseline while showing the persistence of positive coupling to higher redshifts. The black dashed line indicates $\Lambda$CDM ($\beta=0$). Both tests confirm the stability of our results against analysis choices.}
  \label{fig:robustness}
\end{figure*}

{Figure~\ref{fig:robustness} presents robustness tests of our $\beta(z)$ reconstruction for the $\mathrm{CMB}+\mathrm{DESI\,\,DR2}+\mathrm{DESY5}$ combination. The left panel shows the reconstruction using 30 bins, which exhibits the same sign-reversal pattern and similar amplitude of deviations from $\Lambda$CDM as the 20-bin baseline results presented in Fig.~\ref{fig:betaz}. The close agreement between the two demonstrates that our key findings are robust against the specific choice of bin number. The right panel extends the uniform binning range from $z=2.5$ to $z=4.5$. The reconstructed $\beta(z)$ for $z<2.5$ remains fully consistent with our main results, while the positive coupling trend persists to higher redshifts with increasing uncertainties, reflecting the diminishing constraining power of the data. Collectively, these tests validate that the sign-reversal feature and deviations from $\Lambda$CDM reported in our analysis are genuine signals rather than artefacts of specific analysis choices.}

\bibliographystyle{JHEP.bst}
\bibliography{idenpr}

\end{document}